**24-hr Solid-State Power Generation with Self-Adaptive Tunable Radiative Coatings**

*Ken Araki, and Liping Wang[*]*

K. Araki, L. Wang
School for Engineering of Matter, Transport and Energy, Arizona State University, Tempe, USA
[*] Corresponding author. Email: liping.wang@asu.edu

Funding: U.S. National Science Foundation (CBET-2212342)

**Abstract.** Day-night solid-state power generation is experimentally demonstrated with thermoelectric generators integrated with a tunable radiative coating, which switches between a selective solar absorber during daytime and a radiative cooler during nighttime by temperature without any external control. The self-adaptive radiative coating is carefully designed with thermochromic $VO_2$ thin film on heavily doped silicon substrate along with a phase-shifting silicon spacer in between, whose physical mechanism is elucidated. The fabricated coating exhibits a high solar absorptivity of 0.90 and a low infrared emissivity of 0.23 during daytime, and a high emissivity of 0.85 at night, resulting in large emissivity change of 0.62 within the atmospheric transparency window upon metal-to-insulator phase transition. From outdoor tests in a high vacuum apparatus, the tunable radiative coating in 35-mm squared size achieves 1.64 W/m$^2$ power generation around noon time from a stack of commercial thermoelectric modules of 15 mm squared, which is 28% more from that with a black absorber. At night, it produces 26 mW/m$^2$ power in vacuum, outperforming the black emitter by 63%. Even exposed the ambient with convective loss, the 2-inch-round tunable coating generates power of 0.48 W/m$^2$ at noon time, 14% more than the black absorber of the same size. The tunable radiative coating covered thermoelectric modules produces open circuit voltages from 433 mV to −62 mV in vacuum and from 300 mV to −51 mV in ambient through 24-hr continuous outdoor tests.

## 1. Introduction

Photothermal conversion technologies that utilize solar energy and cold outer space have been developed over last decades as one of sustainable solutions to the growing energy demand.[1, 2] Manipulation of light is the key factor to such technology and various materials such as semiconductors, carbon, polymers, and 2D materials have been studied.[3] Spectral selectivity with high solar absorptivity and low infrared emissivity has been recognized as one major focus to enhance the solar thermal conversion long ago for more efficient solar water heaters [4] and recently for solar thermal power generation including solar thermoelectric [5] and solar thermophotovoltaic devices.[6, 7] Various selective solar absorbers have been designed and fabricated that exceed solar absorptivity of 0.8 and emissivity lower than 0.1,[8-12] and stagnation temperature beyond 200°C has been demonstrated in vacuum without optical concentration.[13] On the other hand, spectral selectivity is equally important for radiative sky cooling, as it requires low solar absorption and high infrared emission to achieve sub-ambient daytime temperatures.[14,15] Narrowband emitters which minimize the radiative heating from atmosphere could reach more significant temperature drop in particular under vacuum condition without convection.[16] A variety of materials has been studied for spectrally selective emitters,[17] while LiF as a bulk material is found to be an excellent narrowband emitter with high emissivity of 0.95 within atmospheric transparency window.[18]

Early work utilized near-black solar absorber to heat up the hot side of thermoelectric generators (TEGs) to produce power at daytime but suffered from low performance.[19-23] With selective solar absorbers and high thermal concentration (i.e., absorber-to-TEG area ratio), Kraemer et al. experimentally demonstrated the solar TEG conversion efficiency of 4.6%.[19] Later with segmented thermoelectric legs, high-temperature selective solar absorber, and combined optical and thermal concentration of ~270, they achieved a system efficiency of 7.4% under 38 kW/m$^2$ illumination.[20] On the other hand, sub-ambient temperatures achieved by radiative coolers create negative-temperature differentials across the TEGs, and nighttime solid-state power generation has been experimentally demonstrated very recently. Raman et al. reported power density ~2.5 mW/m$^2$ from single TEG module in ambient during nighttime with a black surface and emitter-to-TEG area ratio ~35.[24] Omair et al. achieved nighttime electricity generation exceeding 100 mW/m$^2$ in ambient from a stack of three TEGs covered by a black emitter with an area ratio ~13.[25] Assawaworrarit et al. demonstrated 350 mW/m$^2$ power generation at night in the vacuum environment with 5 TEG modules cooled by a narrowband selective multilayer emitter with a large area ratio ~280.[26]

It would be incredibly appealing for single TEG device to produce power at both daytime and nighttime or 24-hr continuously. Several works have reported synergetic solar heating and radiative cooling effects to achieve all-day power generation, but most of them involved complicated device structures and achieved small temperature differentials less than 10 K or small power output, whose performances are largely limited by the static properties of materials.[2, 27-33] Tunable materials, whose optical and radiative properties change by voltage or temperature, could be a viable solution. Electrochromic materials are well-studied with variable optical properties mostly in visible and near-infrared by external electrical bias.[34-40] On the other hand, thermochromic materials such as vanadium dioxide ($VO_2$) could experience drastic change in infrared properties upon insulator-metal phase transition around 68°C passively by temperature.[41] While $VO_2$ has been widely used for smart energy-saving windows and passive thermal regulation,[42] very few experimental works reported $VO_2$ based tunable radiative coatings (TRCs) for all-day energy harvesting and power generation.[43-48]

With MBE-grown $VO_2$ thin film coated on sapphire, Ao et al. achieved stagnation temperature rise of 170 K (day) and drop of 20 K (night) in outdoor vacuum tests with infrared emissivity change of 0.5.[43] Araki et al. demonstrated comparable temperature rise of 169 K (day) and drop of 17 K (night) in vacuum with $VO_2$ TRCs fabricated on various substrate materials which exhibit emissivity change up to 0.58.[48] When exposed to ambient, continuous energy conversion is shown to be possible with convective loss, while temperature differential is reduced to 65 K (day) and 5 K (night).[44] By using W-$VO_2$ TRC with emissivity contrast of 0.53, Liu et al. reported ambient TEG power generation of 1 $W/m^2$ during daytime and 25 $mW/m^2$ at night, while the temperature differential across the TEG is less than 3 K.[45] Wu et al. achieved peak power output of 9 $W/m^2$ during the day from the TEG with 34 K temperature differential due to high solar absorptivity 0.96 from the W-$VO_2$ TRC and 20 $mW/m^2$ at nighttime with only 2 K temperature drop because of low emissivity contrast of 0.36.[47]

In this work, we present a self-adaptive $VO_2$ TRC with excellent spectral selectivity at both daytime and nighttime along with large emissivity contrast, to experimentally demonstrate day-night solid-state thermoelectric power generation in both vacuum and ambient conditions. The novel nanophotonic design is articulated with optical modeling with physical mechanism elaborated. The tunable radiative coating up to 2-inch size is successfully fabricated by thin film depositions and low-oxygen thermal oxidation for high-quality $VO_2$. Temperature dependent optical and radiative properties are characterized, and a test apparatus is developed to experimentally demonstrate the day-night thermoelectric power generation with the self-switchable tunable coating both in high vacuum and in ambient. The power generation at both daytime and nighttime with the tunable radiative coating is directly compared to that measured with the black sample. A highly selective metafilm solar absorber and a LiF wafer as a highly selective radiative cooler are also measured to show the importance of spectral selectivity in the TEG power generation. 24-hr continuous power generation is demonstrated in both vacuum and ambient with the tunable radiative coating.

## 2. Results and Discussion

**Figure 1A** depicts the concept of day-night continuous power generation from a stack of TEGs whose top surface is covered by the self-adaptive tunable radiative coating and the bottom side is maintained at ambient temperature. Positive temperature differential across the TEG stack is obtained during daytime with solar heating effect, and negative temperature differential is achieved at nighttime with radiative sky cooling effect. The ideal tunable coating should function as a perfect selective solar absorber ($\alpha_{0.3\text{-}2.5} = 1$ and $\varepsilon = 0$) at daytime and as a perfect selective thermal emitter (emissivity $\varepsilon_{8\text{-}14} = 1$) at nighttime, to maximize the temperature differentials and net heat transfer with property change by temperature. A tunable radiative coating based on $VO_2$ for the self-adaptive energy-harvesting is carefully designed as shown in **Figure 1B**, where it consists of 100-nm $SiO_2$ anti-reflection layer to enhance the solar absorptivity at metallic phase, 300-nm $VO_2$ film for self-adaptive property change, 50-nm $SiO_2$ layer to prevent vanadium diffusion during the oxidation process, and 400-nm sputtered silicon for phase shifting on a 280-μm-thick HDSi substrate ($10^{20}$ $cm^{-3}$) with high infrared absorption. Optical modeling predicts high solar absorption spectrum and low infrared emission with 300-nm $VO_2$ in its metallic phase, which is nearly opaque in the infrared (See **Figure S1** for complex refractive indices). On the other hand, a prominent high emission peak within the atmospheric transparency window (8-14 μm) emerges when $VO_2$ turns insulating, suggesting high emissivity contrast upon phase transition, thanks to the intermediate spacer in combination with HDSi substrate. In comparison, the bare HDSi wafer exhibits high absorption due to free carriers around 6-μm wavelength outside the atmospheric transparency window, let alone tunable radiative property.

To elaborate the design in achieving high emissivity contrast, doping concentration of the HDSi was first optimized to obtain low reflectance with insulating $VO_2$. As shown in **Figure 2A**, doping of $10^{20}$ $cm^{-3}$ leads to the lowest reflectance at 10-μm wavelength, which requires a direct coating of a thick $VO_2$ layer of 1050 nm. Realistically, fabrication of such a thick $VO_2$ is challenging and time consuming. In this work, $VO_2$ thickness is fixed at 300 nm and a high-index intermediate spacer is added between VO2 and HDSi to excite Fabry-Perot like resonance for enhancing absorption in the mid-IR. Different infrared transparent materials such as ZnS ($n$ = 2.2), Si ($n$ = 3.4) and Ge ($n$ = 4.0) are considered, and calculation suggests lowest reflectance around 10-μm wavelength with 400-nm Si, which is chosen over ZnS for much smaller thickness and over Ge for better thermal stability during the vanadium oxidation process at high-temperature. Calculated spectral reflectance of the silicon spacer on HDSi clearly shows the near-zero reflectance dip shifts to longer wavelength with thicker silicon spacer, and finally within the 8-14 μm atmospheric transparency window with 400 nm thickness, as seen in **Figure 2B**. This is further confirmed by the measured reflectance from bare HDSi wafer and that sputtered with 400-nm silicon spacer, which are in excellent agreement with modeling. The underlying physical mechanism can be elucidated by the total phase shift. As shown in **Figure 2C**, 400-nm silicon spacer on HDSi exhibits $\pi/2$ phase shift $\lambda$ = 10 μm, where destructive interference occurs leading to strong absorption. With addition of 400-nm insulating $VO_2$ and 50-nm $SiO_2$, the $\pi/2$ phase shift occurs at $\lambda$ = 11.2 μm still within the atmospheric transparency window. This is confirmed by the calculated reflectance (**Figure 2D**), which reaches zero at $\lambda$ = 10 μm and 11.2 μm for bare Si spacer and $VO_2/SiO_2$ coated Si spacer on HDSi substrate, respectively. Furthermore, the electric field distribution at $\lambda$ = 11.2 μm (**Figure 2E**) clearly reveals the high absorption nature with insulating $VO_2$ and low absorption with metallic $VO_2$ from the final tunable coating structure including the 100-nm $SiO_2$ antireflection layer.

The carefully designed tunable radiative coating was successfully fabricated in two different sizes of 35-mm squared and 2-inch round via thin film deposition processes including sputtering, plasma enhanced chemical vapor deposition, and thermal oxidation, as shown in **Figure 3A**. The temperature-dependent solar absorptivity measurement reveals high solar absorption from 0.4 to 1.6 μm spectral range as predicted with total solar absorptance of 0.90±0.01 with little change at different temperature regardless $VO_2$ phase (**Figure 3B**). However, temperature-dependent infrared spectroscopic measurement unveils the drastic emissivity change upon heating up the tunable radiative coating with $VO_2$ phase transition (**Figure 3C**). In particular, with insulating $VO_2$ at temperatures below 60°C, the tunable coating exhibits a broadband high emission (reaching 100% at $\lambda$ = 11 μm) within atmospheric transparency window (8-14 μm), but it becomes highly reflective upon transition to metallic phase with about 20% emission at temperatures above 80°C. Note that the measured infrared emissivity matches well with modeling in both $VO_2$ phases (see **Figure S2**). By integrating the measured spectral emissivity within the atmospheric transparency window, the total emissivity $\varepsilon_{8\text{-}14}$ is obtained as a function of temperature upon both heating and cooling in **Figure 3D**, where high emissivity of 0.85 with insulating $VO_2$ and low emissivity of 0.23 with metallic $VO_2$ are experimentally achieved. Consequently, a large emissivity contrast of 0.62 upon phase transition is experimentally obtained, which is higher than our previous $VO_2$ tunable coatings fabricated on different substrates[48] and other reported similar $VO_2$ based tunable radiative coatings for day-night energy harvesting and power generation[43-47] (See **Table S1**). The derivative of total emissivity of the tunable radiative coating with respect to temperature shown in **Figure 3E** displays the transition midpoint at 74°C upon heating and 66°C upon cooling along with thermal hysteresis of 8°C.

Several reference samples are used for comparison including a broadband black absorber ($\alpha_s$=0.95, $\varepsilon$=0.99), a highly selective daytime metafilm solar absorber (MetaSA, $\alpha_s$=0.95, $\varepsilon$=0.10),[9] and a highly selective nighttime radiative cooler made of lithium fluoride single-crystal wafer (LiF, $\alpha_s$=0.10, $\varepsilon_{8\text{-}14}$=0.95)[18] (See **Figure S3**). The fabricated $VO_2$ tunable radiative coating (VO2TRC) is placed on a hot plate alongside the three reference samples all in 2-inch size. As shown in **Figure 3F**, visibly the black sample exhibits the darkest color, followed by MetaSA and VO2TRC at room temperature, due to their high solar absorptivity, while LiF appears a much lighter color because of the highly reflective aluminum backcoating. However, under an infrared camera, LiF turns as bright as the black sample because of their close-to-unity infrared emissivity within the 8-14 μm range, while MetaSA appears the darkest owing to its lowest infrared emission. When heated, the reference samples show little color changes under the infrared camera, but the VO2TRC starts to change from bright to dark appearance from 65°C to 95°C, as its emissivity decreases significantly. This confirms the successful suppression of thermal emission upon the insulator-to-metal phase transition, which is crucial to achieve high energy conversion performance during both daytime and nighttime when VO2TRC is paired with TEG modules.

Outdoor power generation performance with fabricated VO2TRCs and reference samples are measured with a home-built test setup, which could provide high vacuum ($P$<0.01Pa) to eliminate convective effect, on the rooftop of Engineering Research Center building at Arizona State University in Tempe AZ, under clear skies during both daytime and nighttime (**Figure 4A)**. Quartz viewport with high solar transmittance (>0.9) is used for daytime vacuum test, while ZnSe viewport is mounted at nighttime for its high infrared transmittance (>0.9) within atmospheric transparency window. Inside the test chamber, 16 TEG modules in 15 mm squared size are stacked on a water block, which maintains the TEG bottom side at ambient temperature. Samples are mounted on the TEG top side with sample-to-TEG area ratio of 5.4 for 35-mm squared samples or 8.4 for 2-inch-round samples.

Daytime vacuum tests for the TEG power generation were conducted in early Nov. 2025 during 6-hr period from 9am – 3pm for the 35-mm VO2TRC and the 35-mm black samples. As shown in **Figure 4B**, the VO2TRC mounted on the TEG top side reaches above 75°C during most of the day, and it functions as the selective solar absorber with fully metallic $VO_2$ phase. This leads to highest temperature of 114°C on the top side of the TEG stack around noon with peak solar irradiance of 928 W/m$^2$ thanks to its high solar absorptivity of 0.90 and low emissivity 0.23. On the other hand, due to its broadband nature, the Black sample could only reach up to 98.4°C under the nearly same solar irradiation (930 W/m$^2$). Note that the bottom side temperature of the TEG stack is about the same during both tests varying between 23°C to 35°C as ambient. With the current-voltage curves are measured at selected hours (**Figure 4C**), the peak power generation is found around noon to be 1.64 W/m$^2$ with the 35-mm VO2TRC sample, which exceeds the peak power density (1.28 W/m$^2$) with the 35-mm black sample by 28%, owing to its excellent spectral selectivity at $VO_2$ metallic phase. Additional daytime vacuum tests show that, a 2-inch near-perfect selective solar absorber (MetaSA, $\alpha_s$=0.95, $\varepsilon$=0.10) could generate peak power of 2.1 W/m$^2$ from the same TEG stack, surpassing that from the 2-inch black sample by 175%. (See **Figure S4**).

Nighttime vacuum tests for the TEG power generation were performed in late Oct. 2025 during 4-hr period from 1am – 5pm for the 35-mm VO2TRC and the 35-mm black samples under relative humidity around 20~30% and ambient temperature around 18°C. As $VO_2$ is in fully insulating phase, the VO2TRC exhibits high infrared emissivity of 0.85 within the atmospheric transparency window, acting as a selective radiative cooler. While the Black sample has about unity emissivity, thanks to the spectral selectivity, the VO2TRC achieved about the same largest negative temperature differential of ~8°C across the TEG stack as the Black one (**Figure 4D**). More importantly, with less radiative heating from the atmosphere outside the transparency window, the 35-mm VO2TRC turns out to produce peak power density of 26 mW/m$^2$ from the TEG stack with 63% improvement compared to 16 mW/m$^2$ from the 35-mm Black sample, as shown in **Figure 4E**. A near-perfect narrowband selective radiative cooler (LiF, $\varepsilon_{8\text{-}14}$=0.95) in 2-inch size could further generate more power of 29

mW/m$^2$, compared to 20 mW/m$^2$ with the 2-inch Black sample from additional nighttime vacuum tests conducted in late June, 2026 (See **Figure S5**).

While vacuum condition is expected to achieve better power generation performance with eliminated convective effect, it would unavoidably make the device more complicated and less practical. By using the same test setup without vacuum, daytime tests under ambient condition were conducted during 6-hr period from 9am – 3pm for the 2-inch VO2TRC and 2-inch Black samples in late July, 2026 under extreme heat climate with ambient temperatures varying from 33°C to 50°C. As shown in **Figure 5A,** the 2-inch VO2TRC sample at the TEG top side stays in its fully metallic phase as selective solar absorber at temperatures above 75°C after 10:30am with a peak temperature 87°C reached around noon under maximum solar irradiance 961 W/m$^2$. On the other hand, the 2-inch Black sample only reaches a maximum of 67°C during the day. The 20°C more temperature rise on the TEG top side is a result of excellent spectral selectivity of the VO2TRC as selective solar absorber with $VO_2$ metallic phase. However, the temperature difference across the TEG stack is 38.5 K for the VO2TRC and 23.5 K for the Black sample around noon because of higher ambient temperature by 5°C on the day that VO2TRC is tested. Nevertheless, the TEG stack covered by the 2-inch VO2TRC sample produces 0.48 W/m$^2$, slightly more than 0.42 W/m$^2$ from that with the 2-inch Black sample. Clearly, the daytime TEG power generation can be significantly degraded by the convection heat loss as well as higher ambient temperature that decreases the temperature differential across the TEG stack.

To demonstrate the convection effect on the nighttime power generation performance, the 2-inch VO2TRC sample was tested under both vacuum and ambient conditions during 4-hr period from 1am – 5pm in early July, 2026 with relative humidity around 10~20%. As shown in **Figure 5C**, the VO2TRC leads to about the same 10°C negative temperature differential across the TEG stack under both the vacuum and ambient conditions, while the TEG bottom temperature is about 2°C higher when tested in vacuum than that tested in ambient. Without convective heating, the 2-inch VO2TRC covered TEG stack generates 20 mW/m$^2$ in vacuum with 43% better performance than that (14 mW/m$^2$) in ambient condition. Also, while higher nighttime power generation is expected with larger sample size, however, the 2-inch VO2TRC produces slightly less power than the 35-mm VO2TRC both in vacuum, which is attributed to more radiative heating from the atmosphere and vacuum chamber wall at higher ambient temperature by about 10°C.

Finally, 24-hr continuous power generation was tested in vacuum with the 35-mm VO2TRC sample on Nov. 9, 2025 (6am – 6am) and in ambient with the 2-inch VO2TRC sample on July 23, 2026 (6:30pm – 6:30pm), where the measured temperatures across the TEG stack and the open-circuit voltage ($V_{oc}$) are shown in **Figure 6A** and **6B**, respectively. With the VO2TRC sample in vacuum, the TEG stack starts to generate positive $V_{oc}$ from positive temperature differential $\Delta T$ around 7:30am, and reaches a maximal $V_{oc}$ of 433 mV with highest TEG top side temperature ~110°C and largest $\Delta T$ of 74 K under peak solar irradiance 920 W/m$^2$ around noon. After sunset around 5:30pm, negative $V_{oc}$ is produced with −62 mV and negative temperature differential $\Delta T$ of −10.4 K for most of the nighttime. With the 2-inch VO2TRC exposed to hot ambient with the TEG bottom side temperature about 15°C higher than that tested in vacuum, the TEG stack starts to produce positive $V_{oc}$ around 7:30am after the sunrise, and reaches a smaller maximal $V_{oc}$ of 300 mV with highest TEG top side temperature 86°C and $\Delta T$ of 37 K under peak solar irradiance 947 W/m$^2$ around noon. Around 6:30pm before the sunset, negative $V_{oc}$ is generated with −51 mV and negative temperature differential $\Delta T$ of −6 K for most of the nighttime.

## 3. Conclusion

This work has successfully demonstrated a novel $VO_2$ tunable radiative coating with highest emissivity change of 0.62 reported so far with scalable fabrication process (see **Table S1**) for day-night solid-state power generation from the hotness of the Sun and the coldness of the universe. The daytime and nighttime outdoor tests in both vacuum and ambient conditions undoubtedly confirm the superior power generation performance than the black sample when paired with TEG stacks due to its excellent spectral selectivity and self-adaptive tunability. Measurements with near-perfect selective solar absorber (MetaSA) and selective radiative cooler (LiF) suggest room for performance enhancement by larger emissivity contrast and improved spectral selectivity. In addition, it is crucial to carefully minimize convective and radiative heat loads for larger temperature differentials, greater Voc and higher power generation. The device conversion performance can be also optimized with larger sample-to-TEG area ratio as well as TEG module parameters such as Seebeck coefficient or ZT value, electrical resistance and thermal resistance, and number of stacks through rigorous opto-thermal-electrical modeling.

## 4. Methods

*Optical modeling*: Multilayer thin film optics based on transfer-matrix formulation is used to calculate the solar and infrared emittance of the tunable radiative coatings.[49] Optical constants of $SiO_2$ and sputtered Si are taken from Palik.[50] Dielectric function of the heavily doped silicon follows a Drude model.[51] Complex refractive indices of the $VO_2$ film are obtained from fitting the measured reflectance and transmittance spectra in its insulating and metallic phases.[52] The total phase shift is calculated as,[53]

$$\phi = \left(2\beta_2 + 2\beta_1 - \arg(r_{21}) + arg(r_{234}) + arg(t_{21}t_{12} - r_{21}r_{12})\right)/2$$

where $\beta_j$ is the phase shift due to Si ($j$ = 2) and $VO_2$ ($j$ = 1) layer calculated with wavelength $\lambda$, permittivity $\varepsilon_j$ and incident angle $\theta_j$ by,

$$\beta_j = 2\pi Re\left(\sqrt{\varepsilon_j}\right) cos\theta_j/\lambda$$

Here, $r_{ij}$ and $t_{ij}$ are Fresneal coefficients calculated for both forward and backward waves. The reflectance coefficient $r_{234}$ is for 3-layer structure (Si – HDSi – Al) is written based on the Airy formula as,

$$r_{234} = r_{23} + \frac{t_{23}t_{32}r_{34}\exp(2i\varphi)}{1 - r_{32}r_{34}\exp(2i\varphi)}$$

*Sample fabrication*: Highly doped silicon wafer (double-side polished, 280-µm thick) obtained from UniversityWafer with $10^{20}$ $cm^{-3}$ doping level is carefully cleaned with acetone and IPA to remove any contaminations. First, 400-nm silicon is first deposited via RF sputtering at 0.5 Å/s, followed by 50-nm $SiO_2$ diffusion barrier layer deposited via plasma enhanced chemical vapor deposition (PECVD). Then 150-nm vanadium is deposited via DC sputtering at 0.9 Å/s and then oxidized in low-oxygen furnace at 500°C with optimized nitrogen flow for a total of 12 hours.[54] Lastly 100-nm $SiO_2$ as an antireflection coating is deposited via PECVD. The highly selective metafilm solar absorber is fabricated on 2-inch-round polished silicon wafer following the procedures in our previous work.[9] The Metal Velvet black coating on foils purchased from Actar is used as the black absorber after cutting into different sizes. The 2-inch-round single-crystal LiF wafer (double-side polished, 500-µm thick) acquired from Stanford Advanced Materials is used as highly selective radiative cooler.[18] 200-nm aluminum is sputtered on the backside of all the samples to minimize thermal radiation loss.

*Spectral measurements*: Temperature-dependent spectral absorptivity within the solar spectrum is measured with a tunable light source (Newport, TLS-250Q) and an 8-inch PTFE integrating sphere, while temperature-dependent spectral infrared emissivity is measured with a Fourier-transform spectrometer (Thermo Scientific, Nicolet iS50) and a reflection accessory (Harrick Scientific, Seagull). Samples are heated with home-made temperature stages made of thermoelectric module, water block and PID controller. Both solar and infrared spectra are measured at every 5°C from 25°C to 60°C for the insulating phase, every 2°C from 60°C to 80°C within the phase transition, and every 5°C from 60°C to 95°C for the metallic phase upon both heating and cooling.

*Outdoor measurements*: Samples sit on the top surface of a 16-TEG stack (DigiKey, MGM250-31-10-16) inside the vacuum chamber with about 5 mm spacing to the viewport. All the interfaces are filled with thermal paste to minimize contact thermal resistance. The outdoor tests in vacuum use Quartz and ZnSe viewports respectively during daytime and nighttime, and tests start after 1-hr pumping until the pressure reaches below 0.01 Pa. Ambient tests are conducted with the same setup by replacing viewports with thin polyethylene films for shielding the dust and wind. The entire setup is tilted by 20° towards south only at daytime. The sample temperature is measured with a thermistor (Amphenol, SC30F103V) during daytime and by an infrared camera (FLIR, E8-XT) after careful calibration at nighttime.[48] The bottom side of TEG stacks is maintained at ambient temperature by sitting on a water block with running water to a reservoir in thermal equilibrium with ambient. The temperatures of TEG stack bottom side, water block and vacuum chamber wall are measured by thermistors with difference less than 2°C. Current-voltage curves from the TEG stack are collected with a source meter (Keithley 2400), and the TEG power density is calculated as the product of voltage and current normalized to the radiative coating surface area. Solar irradiance is measured by a pyranometer placed on the same setup next to the vacuum chamber, and relative humidity at nighttime is monitored by a weather station.

**Supporting Information**

Supporting figures S1 – S5 and table S1 are included.

**Acknowledgements**

This work was supported by the U.S. National Science Foundation under Grant No. CBET-2212342. We would like to thank ASU NanoFab for use of their nanofabrication and characterization facilities.

**Conflict of Interests**

The authors have no conflict of interest to declare.

**Data Availability**

The data to support the findings of this work is available upon reasonable request.

Author Contributions

K.A. conducted optical modeling, fabricated samples, performed optical characterizations, carried out outdoor tests, analyzed data, prepared figures, and wrote manuscript draft; L.W. conceived the idea, supervised the project, secured the funding, provided computational resources and lab facilities, and revised figures and manuscript. All authors reviewed and approved the final manuscript.

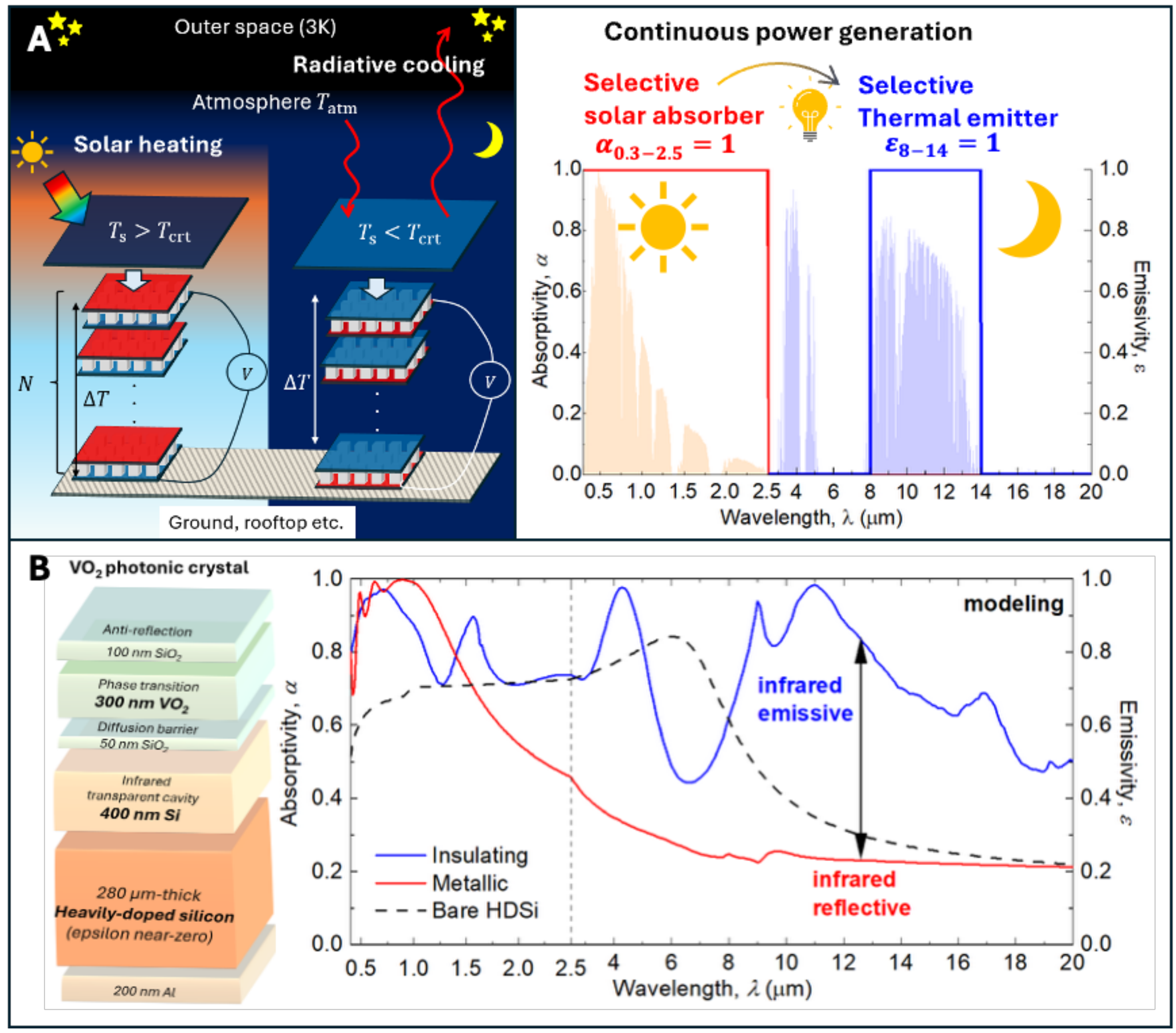

**Figure 1. A.** Illustration of day-night solid-state power generation with a tunable radiative coating on stacked TEGs exposed to the hot Sun for daytime solar heating and the cold Universe for nighttime radiative sky cooling. The tunable radiative coating behaves as a selective absorber during daytime and as a selective thermal emitter during nighttime to achieve high power generation performance. **B.** Design of proposed tunable radiative cooling made of thermochromic $VO_2$ thin film with $SiO_2$ antireflection layer, $SiO_2$ diffusion barrier layer phase-shifting silicon spacer layer on heavily doped silicon wafer (HDSi). The expected solar absorptivity and infrared emissivity spectra are calculated for both insulating and metallic $VO_2$ phases along that of bare HDSi.

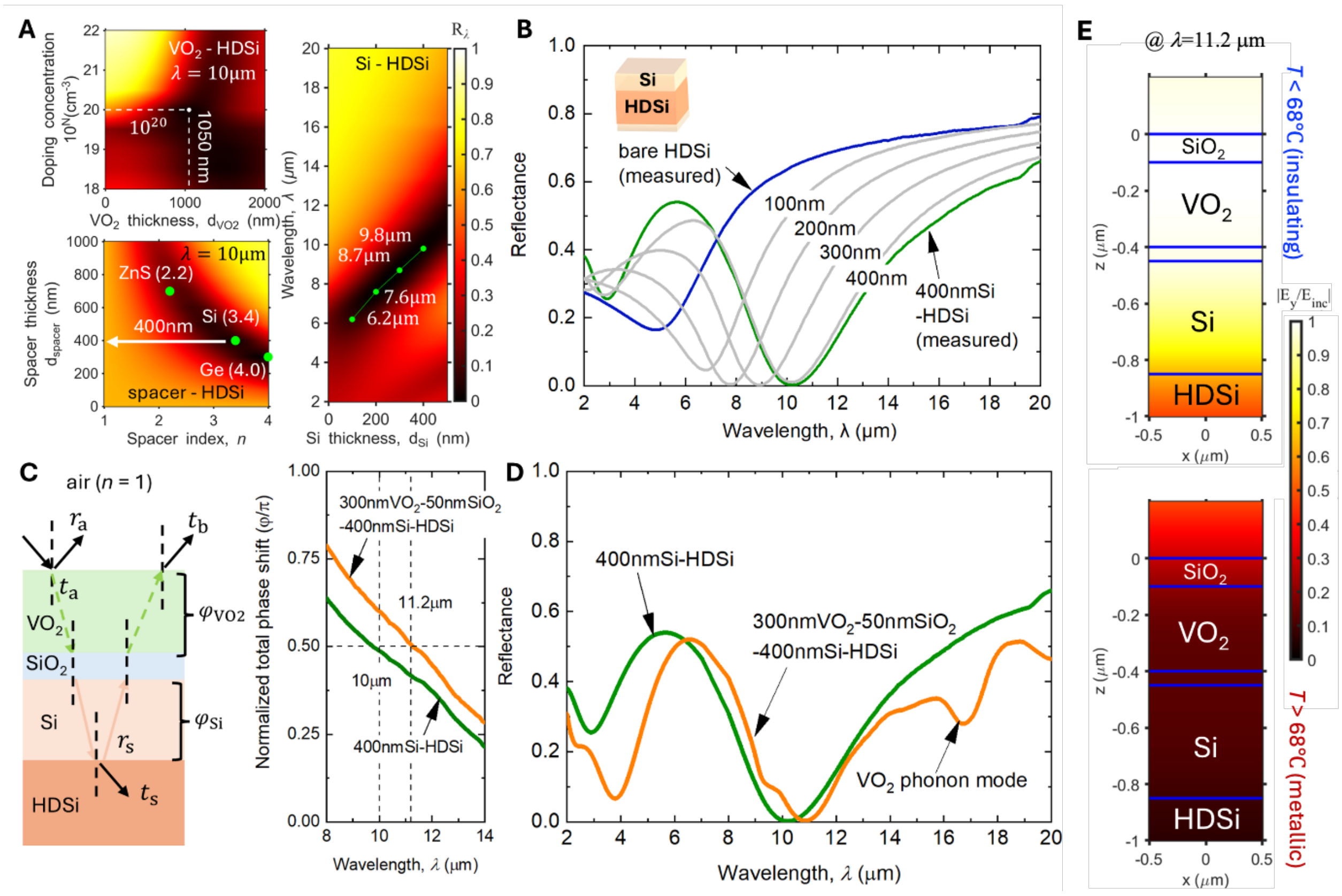

**Figure 2. A.** Effects of HDSi doping concentration, spacer index, and spacer thickness on the reflectance at wavelength $\lambda$=10 μm for the proposed tunable radiative coating with insulating $VO_2$ from modeling. **B.** Calculated reflectance spectra of the sputtered silicon spacer layer HDSi with different spacer thickness along with measurements. **C.** Schematic of wave propagation through $VO_2$/$SiO_2$/Si multilayer on the HDSi substrate and calculated total phase shift in the insulating phase. **D.** Comparison of reflectance spectra with and without top 300-nm-thick insulating $VO_2$ layer from calculation. **E.** Simulated electric field distribution at wavelength $\lambda$=11.2 μm within the proposed tunable radiative coating at both insulating and metallic $VO_2$ phases.

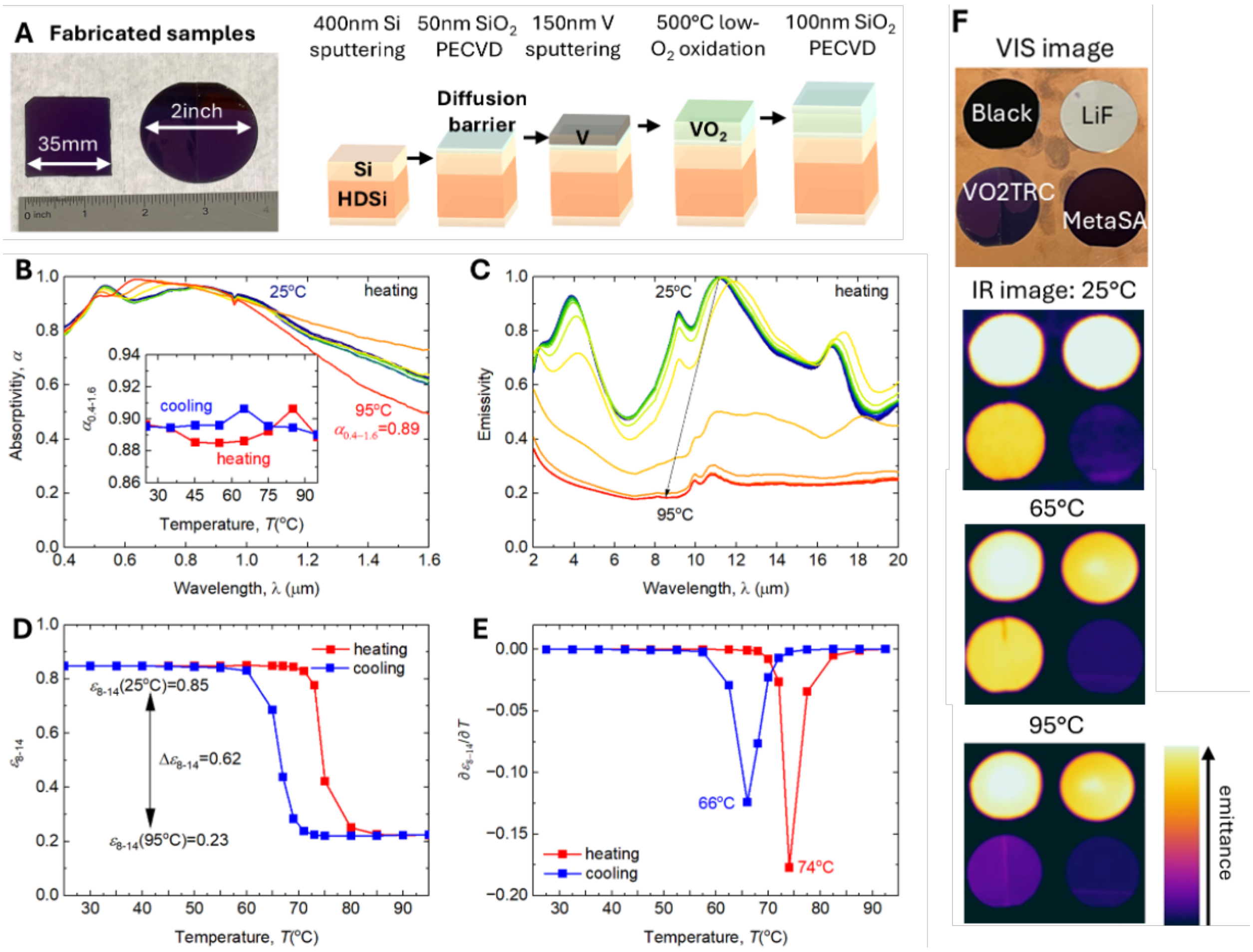

**Figure 3. A.** Photo of fabricated $VO_2$ based tunable radiative coating (VO2TRC) samples in different sizes and fabrication process flow. **B.** Measured temperature-dependent solar absorptivity, **C.** infrared emissivity spectra, **D.** heating-cooling curves and **E.** its derivative of infrared emissivity $\varepsilon_{8\text{-}14}$ of fabricated VO2TRC samples. **F.** Visible and infrared thermographic images of 2-inch Black, VO2TRC, LiF and MetaSA samples placed on hot plate heated at 25°C, 65°C, and 95°C.

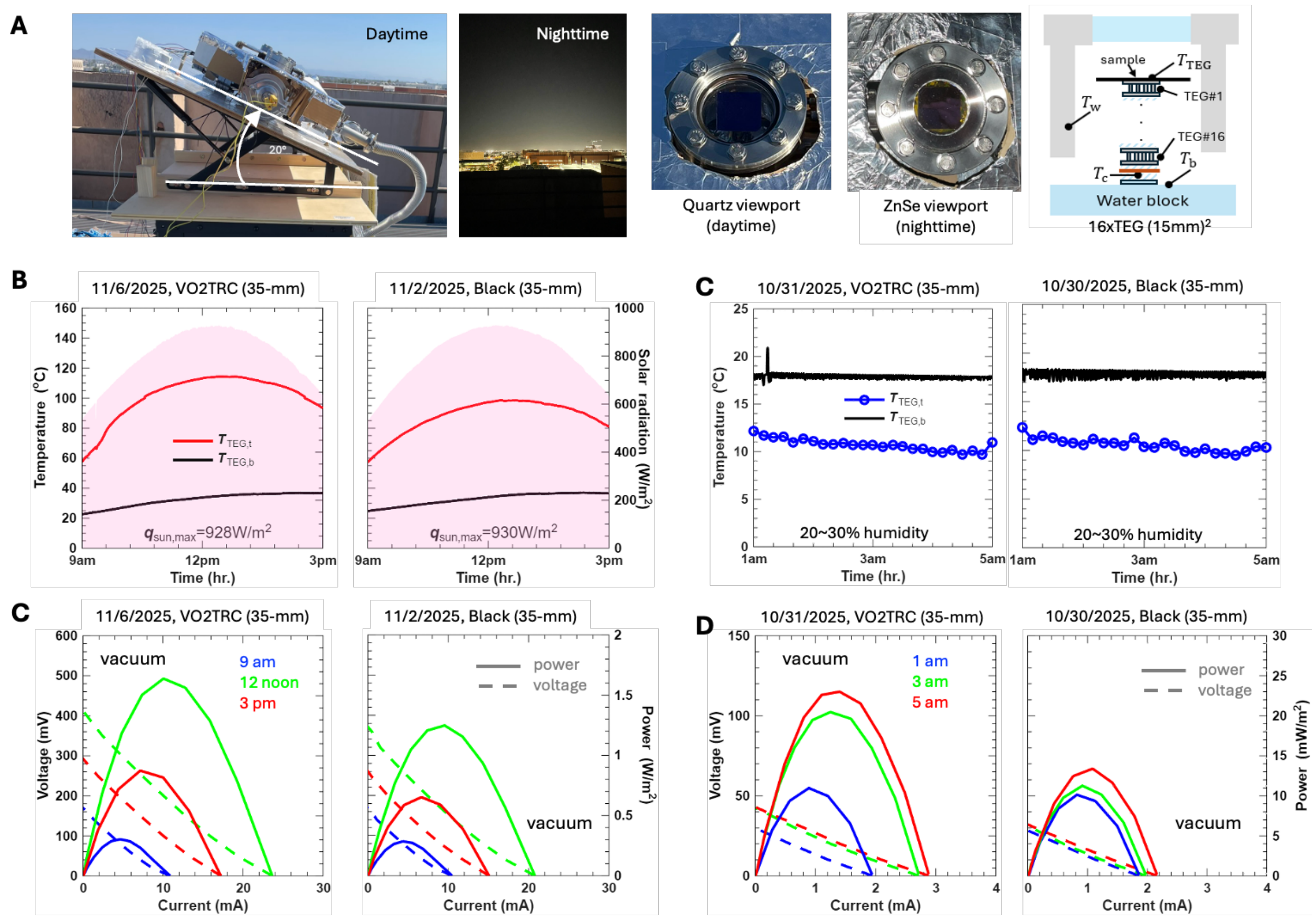

**Figure 4. A.** Photos of the outdoor vacuum test setup on the building rooftop at Tempe, AZ and clear sky during daytime and nighttime. The vacuum setup is tilted at angle of 20° facing South. Quartz and ZnSe viewports are used during daytime and nighttime, respectively. A total of 16 TEGs in 15-mm-squred size is stacked and connected in-series with radiative coating samples attached on top. **B.** Measured temperatures at the top and bottom of the TEG stack and **C.** current-voltage curves and generated power density during daytime in vacuum for the VO2TRC sample and the black sample in early Nov. 2025. **D.** Measured temperatures at the top and bottom of the TEG stack and **E.** current-voltage curves and generated power density during nighttime in vacuum for the 35-mm VO2TRC sample and the 35-mm black sample in late Oct. 2025. Both the VO2TRC and the black samples are in 35-mm squared size.

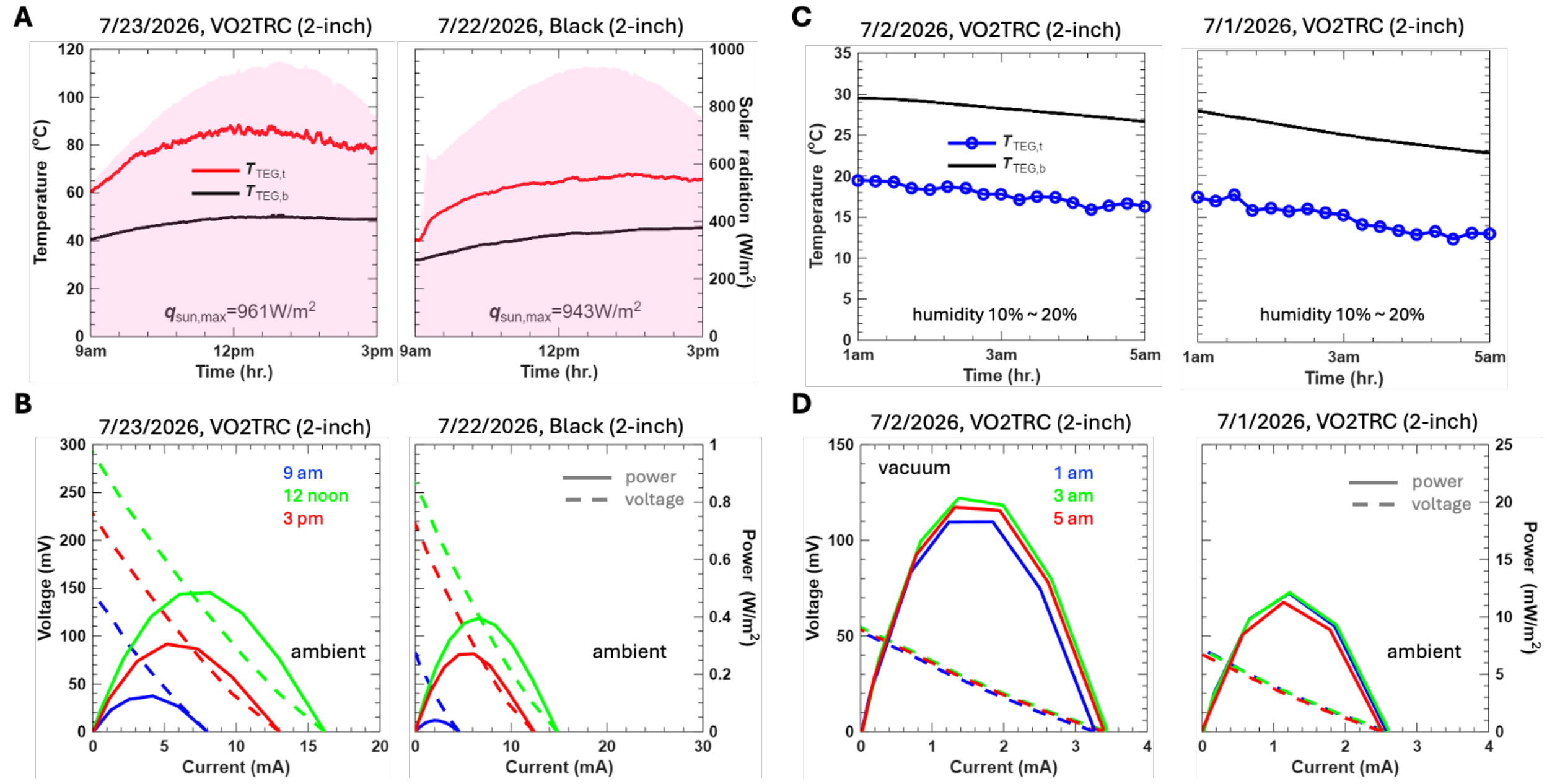

**Figure 5. A.** Measured temperatures at the top and bottom of the TEG stack and **B.** current-voltage curves and generated power density during daytime in ambient for the 2-inch VO2TRC sample and the 2-inch black sample in late July 2026. **C.** Measured temperatures at the top and bottom of the TEG stack and **D.** current-voltage curves and generated power density during nighttime for the 2-inch VO2TRC sample in vacuum and in ambient around early July 2026.

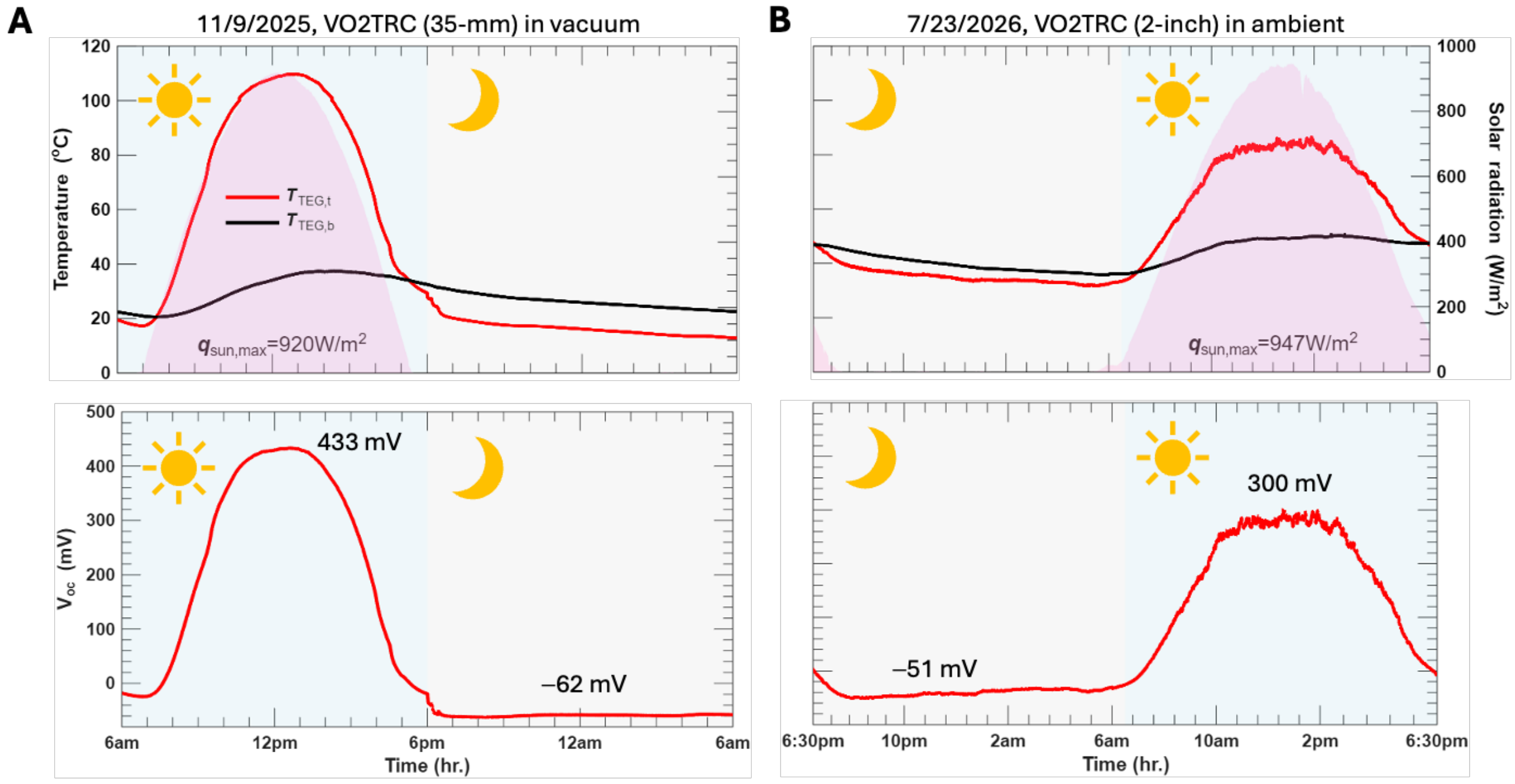

**Figure 6.** 24-hr continuous measurements of temperatures at the top and bottom of the TEG stack and open-circuit voltage during daytime and nighttime with **A.** the 35-mm VO2TRC sample in vacuum on Nov. 9, 2025, and **B.** the 2-inch VO2TRC sample in ambient on July 23, 2026.